%% file: earnings_alignment.tex
\documentclass[sigconf]{acmart} 
\usepackage{graphicx}
\usepackage{subcaption}
\usepackage{hyperref}
\usepackage{url}
\usepackage{algorithm}
\usepackage{algorithmic}

\usepackage{amsmath, amsfonts}
\usepackage{bbold}

\newcommand{\calA}{\mathcal{A}}

\newcommand{\bs}[1]{\left\{ #1\right\}}

\usepackage[textsize=tiny,textwidth=1.4cm]{todonotes}

\AtBeginDocument{%
  }

\copyrightyear{2026} 
\acmYear{2026} 
\acmConference[Preprint]{ }{ }{ }
\acmBooktitle{Preprint}
\acmDOI{10.xxxx/xxxxxxxx.xxxxxxx}
\acmISBN{xxxxx}

\begin{document}

\title{Measuring What is Repeated and Novel in Earnings Disclosures Using Optimal Transport}

\author{Yuntao Wu}
\email{winstonyt.wu@mail.utoronto.ca}
\orcid{0009-0006-6269-1603}
\affiliation{%
  \institution{University of Toronto}
  \city{Toronto}
  \state{Ontario}
  \country{Canada}
}
\author{Lynn Tao}
\email{lynnyl.tao@mail.utoronto.ca}
\orcid{0009-0000-3856-2146}
\affiliation{%
  \institution{University of Toronto}
  \city{Toronto}
  \state{Ontario}
  \country{Canada}
}
\author{Charles Martineau}
\email{charles.martineau@rotman.utoronto.ca}
\orcid{0000-0002-6896-184X}
\affiliation{%
  \institution{University of Toronto}
  \city{Toronto}
  \state{Ontario}
  \country{Canada}
}
\author{Vincent Grégoire}
\email{vincent.gregoire@hec.ca}
\orcid{0000-0002-1230-5026}
\affiliation{%
  \institution{HEC Montréal}
  \city{Montréal}
  \state{Québec}
  \country{Canada}
}
\author{Andreas Veneris}
\email{veneris@eecg.toronto.edu}
\orcid{0000-0002-6309-8821}
\affiliation{%
  \institution{University of Toronto}
  \city{Toronto}
  \state{Ontario}
  \country{Canada}
}

\renewcommand{\shortauthors}{Y. Wu, L. Tao, C. Martineau, V. Grégoire, A. Veneris}

\begin{abstract}

We introduce an optimal transport framework to decompose the textual content of earnings press releases and conference calls into aligned (shared) and unaligned (unique) components, and use these as predictors of stock returns around earnings announcements. Using over 105,000 document pairs from 2008 to 2023, we find that the aligned portions of both disclosures explain announcement-day returns with high statistical significance. The unaligned component of conference calls, which captures forward-looking content absent from press releases, provides substantial incremental explanatory power, shedding light on why conference calls are central to market reactions to earnings news. Extending to post-announcement returns, investors appear to overreact to shared press release content while underreacting to the unique information conveyed in conference calls.

\end{abstract}

\ccsdesc[500]{Computing methodologies~Natural language processing}
\ccsdesc[500]{Applied computing~Economics}

\keywords{alignments, BERT, earnings announcements, market efficiency, natural language processing, optimal transport, topic modeling}


\maketitle

\section{Introduction}

Earnings announcements are among the most important recurring news events in equity markets. Each quarter, firms communicate with investors through two distinct but complementary disclosures: a written press release and a live conference call with financial analysts. A large literature documents that prices respond not only to quantitative earnings surprises but also to the qualitative, textual content of these disclosures \cite{akey2022price,earnings1, price2012earnings, meursault2023pead, garcia2023colour}. 
Conference calls frequently revisit content already contained in the press release, yet these two disclosure channels are rarely studied jointly. A fundamental question remains open: to what extent do press releases and conference calls convey the same information, and does the distinction between shared and novel content help explain how markets react to earnings news?


We examine this question using optimal transport, a mathematical framework for measuring similarity between distributions while accounting for their structure. We apply this approach to paragraph-level alignment of earnings disclosures, allowing us to identify which parts of a press release and conference call convey shared information and which parts contain source-specific content. For each of over 105,000 firm-quarter observations spanning 2008 to 2023, we classify every press release and conference call paragraph as either aligned (semantically shared) or unaligned (unique to its source), and construct return forecasts from each component using FinBERT embeddings and a rolling Ridge regression.

Several findings emerge. 
Both press releases and conference calls individually predict announcement-day abnormal returns and, as measured by Shapley values, both outperform earnings surprise. Once we apply the decomposition, the unaligned component of conference calls emerges as an informative signal to explain announcement-day returns in addition to the shared information: forward-looking content discussed during analyst questioning provides substantial incremental explanatory power. In comparison, press-release content not revisited on the call is statistically uninformative for prices. This sheds light on why the Q\&A session, rather than the prepared remarks in conference calls, drives the share of price formation at the time of earnings announcements \citep{matsumoto2011makes, fan2025investors}. 
A triple sort on earnings surprises and qualitative information from press releases and conference calls shows substantial cross-sectional variation in announcement-day returns, with the strongest spread occurring when earnings surprises and unaligned conference-call content move jointly from the lowest to highest quintiles (-5.6\% to 5.5\%), highlighting the incremental value of novel information revealed during conference calls.

Turning to post-announcement returns, we document asymmetric long-horizon reactions to information conveyed through press releases and conference calls. Over a three-month horizon, stock returns correct the initial adjustments on the announcement day from press releases (i.e. price reversal), especially among microcap firms, consistent with \citet{huang2018headline}. Meanwhile, stock returns continue to drift in the direction implied by the unaligned component of conference calls, indicating persistent underreaction to novel information revealed during the call. Most importantly, this predictability remains statistically significant even among large, liquid firms for which aggregate conference-call measures exhibit no long-run forecasting power, revealing a signal that is entirely obscured without the aligned/unaligned decomposition. Topic analysis using online Latent Dirichlet Allocation with GPT-4o labeling confirms the structural distinction: press-release-specific content concentrates on backward-looking historical performance, whereas conference-call-specific content covers growth opportunities, cost improvements, and forward-looking managerial guidance. 

\section{Related Work}

Prior research in finance examines how investor reactions and prices are affected by earnings surprises (hard information) and the qualitative content (soft information) embedded in the two earnings disclosures. 
For press releases, \citet{akey2022price} show that hackers who stole filings ahead of announcements traded on both earnings surprises and the textual content of those releases, and \citet{earnings1} demonstrate that embedding-based measures of press release content predict announcement-day returns. For conference calls, \citet{price2012earnings}, \citet{meursault2023pead}, \citet{garcia2023colour}, and \citet{ecc-analyzer} document that the language used by managers and analysts is meaningfully related to how stock prices respond on the day of the earnings news. 
Meanwhile, earnings surprises, have become increasingly less predictive of post-announcement returns as markets have become more efficient at processing hard information \citep{martineau2021rest}.

A limitation of existing approaches is that they typically treat each disclosure channel as an independent information source. However, conference calls frequently revisit material already disclosed in press releases.
Thus, using raw conference call content as a predictor may conflate genuinely new information with material investors have already processed. For example, \citet{didisheim2026inefficient} show that stripping out stale, anticipated content in the news media writing about firms and focusing on the residual, the ``unexpected'' portion of the news, produces substantially stronger return predictions. Their findings would imply that in our setting, the predictive power of news embedded in conference calls may be concentrated in content that goes beyond what the press release has already disclosed.

We build on this insight by using optimal transport to perform the decomposition directly at the document level, identifying which paragraphs in the press release and conference call are semantically aligned (shared) and which are unaligned (unique to each source). Optimal transport \citep{monge1781, kantorovich1942, Santambrogio2015} constructs a globally optimal matching between distributions by jointly considering all pairwise costs, and has been applied in textual matching for retrieval and translation. 
\citet{swanson2020} focused on constrained text matching for text selection and \citet{ot-alignment} applied it to align words across languages for cross-lingual transfer learning. We extend this approach from word-level to paragraph-level alignment, which is better suited to the earnings disclosure setting: press releases and conference call transcripts are long documents in which stopwords and redundant phrases can dilute word-level embeddings, whereas paragraph-level alignment captures the semantic structure of the argument being made. This allows us to construct aligned and unaligned soft-information measures that isolate what is genuinely new in each disclosure channel.


\section{Data}

Our sample period spans January 2008 to December 2023. Earnings dates, times, analyst forecasts, and announced EPS are retrieved from LSEG IBES Academic. Daily stock returns and prices are retrieved from CRSP Daily Stock Files. Earnings press releases and conference call transcripts are retrieved from EDGAR and S\&P Global, respectively.

Following \citet{martineau2021rest}, we restrict the sample to earnings announcements with analyst coverage, which allows us to construct earnings surprise as
$$\text{Surprise}_{c,\tau} = \frac{\text{EPS}_{c,\tau} - \mathbb{E}_{\tau-1}[\text{EPS}_{c,\tau}]}{P_{c,\tau-5}},$$
where $\text{EPS}_{c,\tau}$ is firm $c$'s reported earnings per share for the quarterly announcement on date $\tau$, and $\mathbb{E}_{\tau-1}[\text{EPS}_{c,\tau}]$ is the consensus analyst forecast, defined as the median of all forecasts issued in the 90 days prior to the announcement. We scale by the stock price $P_{c,\tau-5}$ measured five trading days before the announcement and winsorize at the 1st and 99th percentiles within each calendar year.

We retrieve earnings press releases (PR) in HTML format from 8-K filings on SEC EDGAR. We retain only releases published outside of regular trading hours — on or after 4:00 p.m. or strictly before 9:30 a.m. ET.
This yields 132,352 press releases covering 5,773 unique firms (identified by PERMNO in CRSP). Announcement-day returns are computed as the percentage change in stock price from the prior close to the post-announcement close. We truncate each press release to its first 20 paragraphs and remove boilerplate sections, leaving approximately 10,000 characters per document following \citet{akey2022price}. 

We retrieve earnings conference call transcripts from S\&P Global and use only the Q\&A component (CC), setting aside the prepared-remarks presentation. This choice is standard in the literature: the Q\&A session is the primary driver of the price variation associated with conference calls, while the presentation adds comparatively little incremental information \citep{matsumoto2011makes, fan2025investors}. Q\&A transcripts contain approximately 20,000 characters on average.

\begin{figure*}[htb]
  \centering
  \begin{subfigure}[b]{0.3\linewidth}
  \centering
  \includegraphics[width=\linewidth]{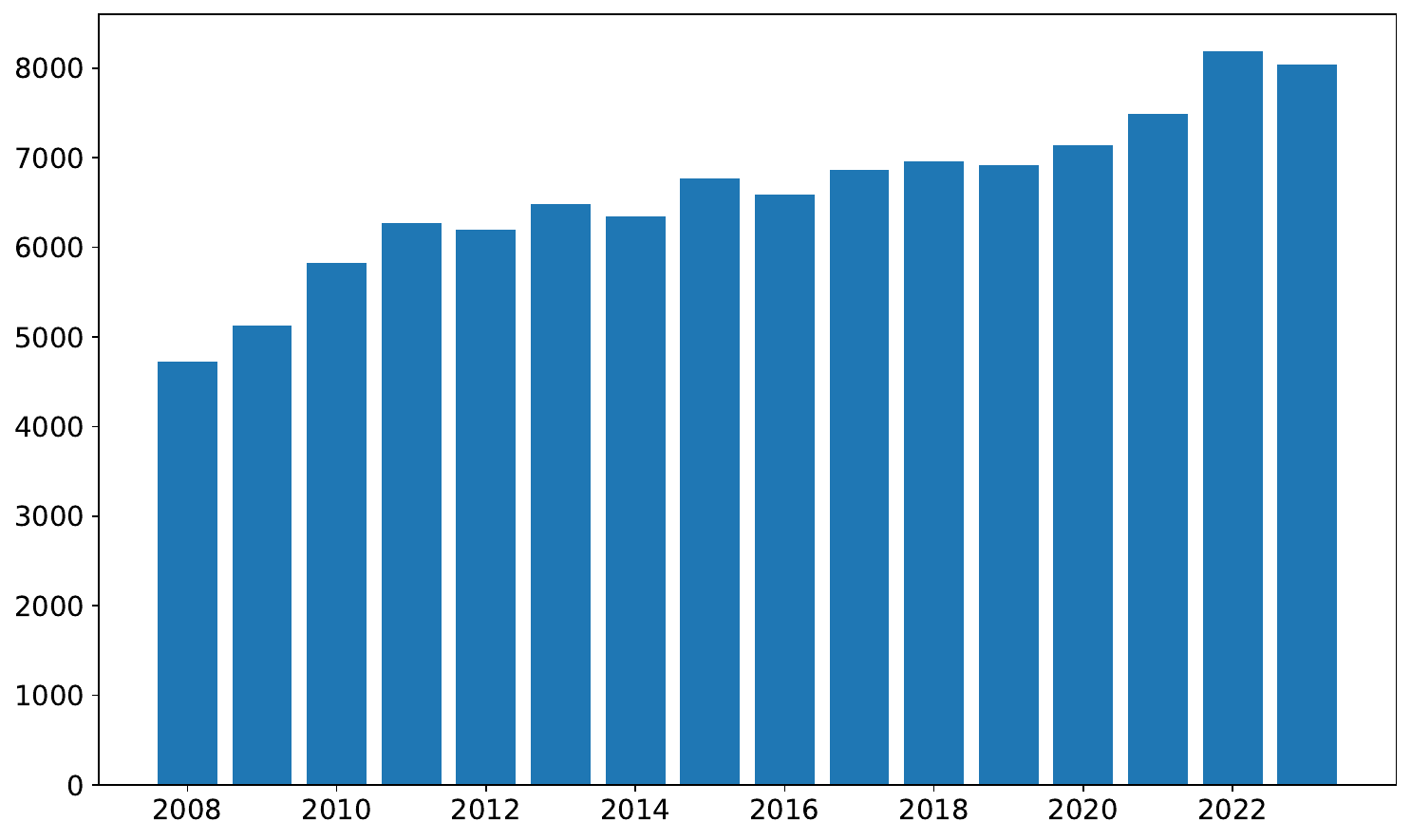}
  \caption{Merged Calls}
  \end{subfigure}
  \hfill
  \begin{subfigure}[b]{0.3\linewidth}
  \centering
  \includegraphics[width=\linewidth]{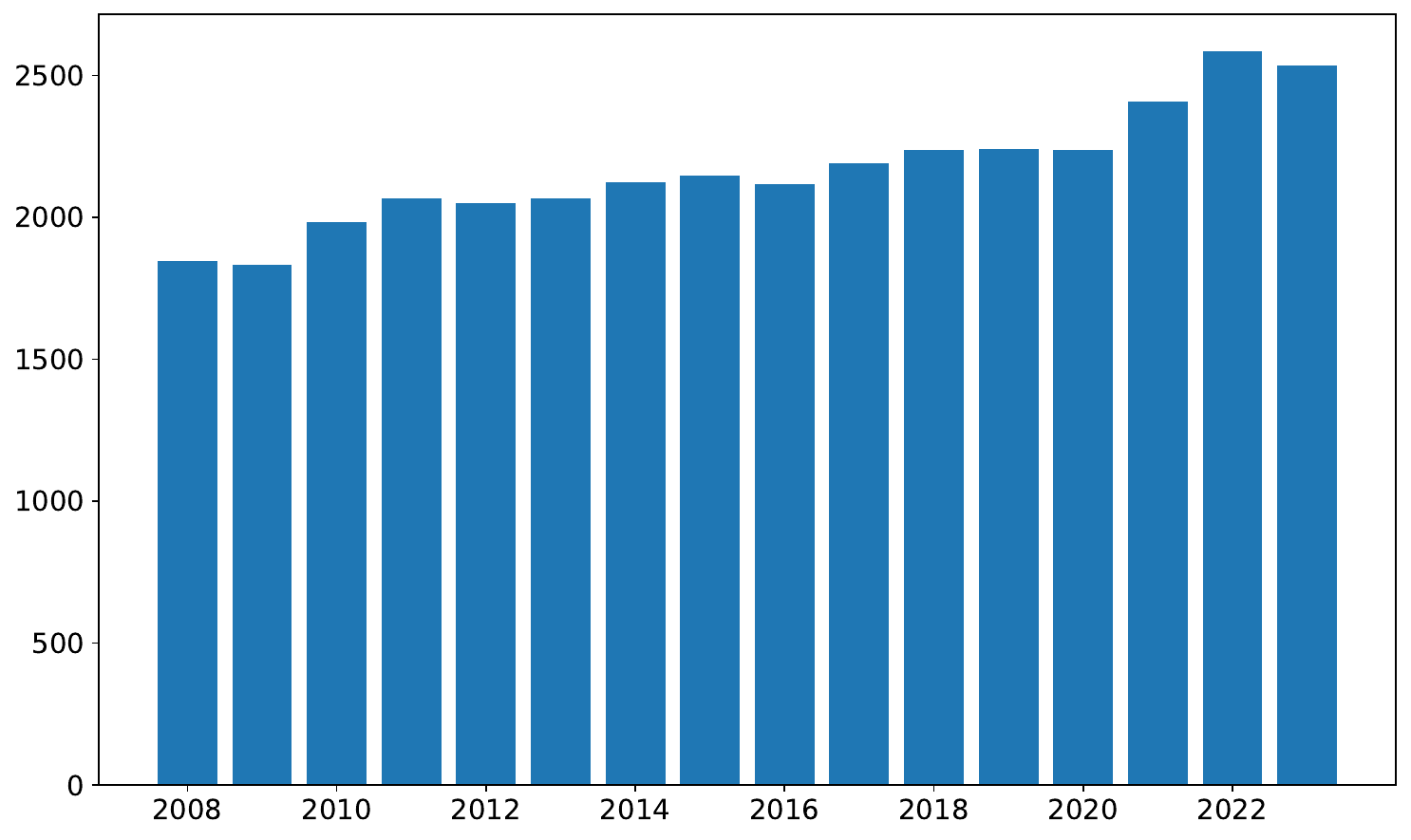}
  \caption{Merged Stocks}
  \end{subfigure}
  \hfill
  \begin{subfigure}[b]{0.3\linewidth}
  \centering
  \includegraphics[width=\linewidth]{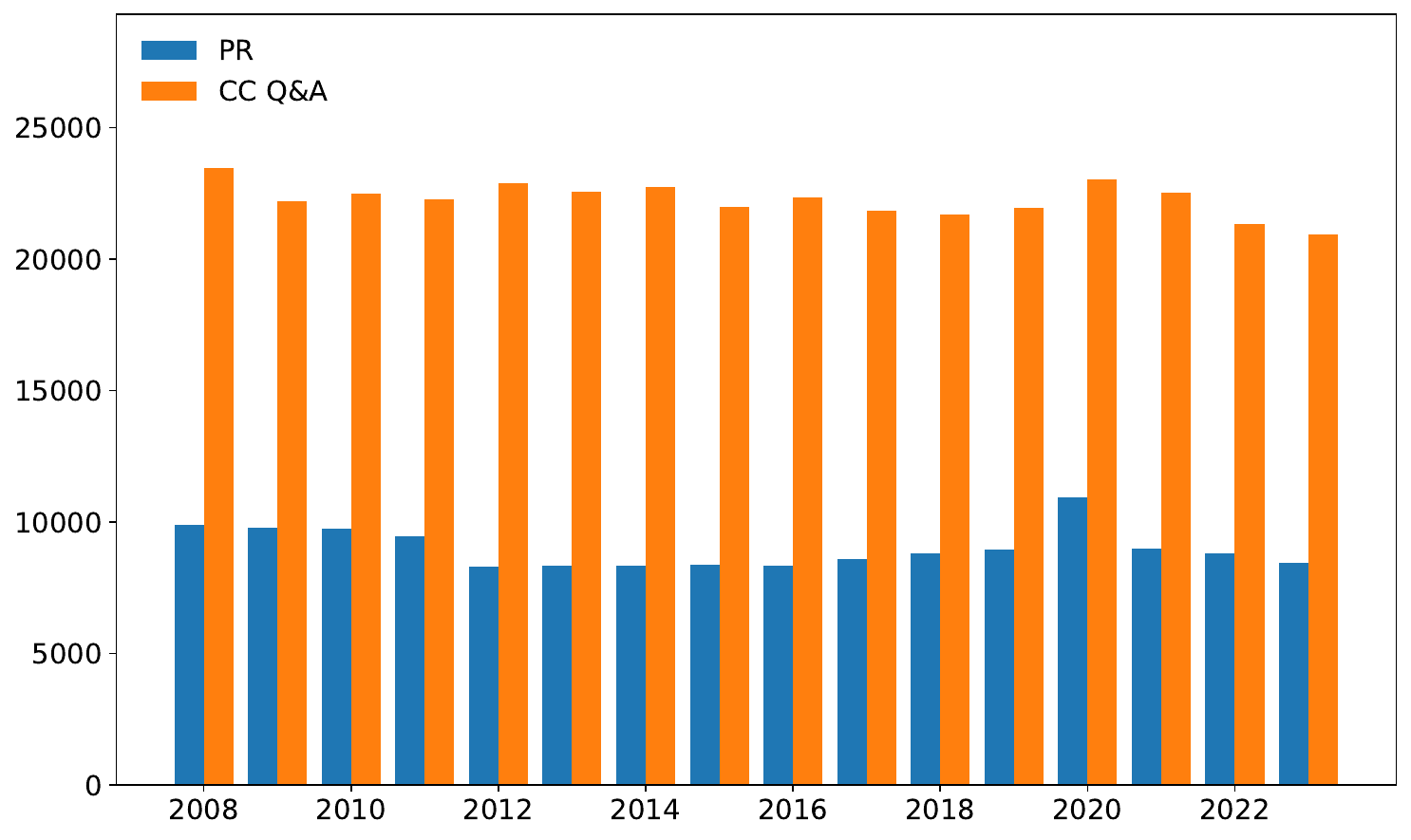}
  \caption{Character Counts}
  \end{subfigure}
  \caption{General Statistics \label{fig:statistics}}
  \Description{General Statistics}
\end{figure*}

We merge press releases, conference call transcripts, and returns using PERMNOs and GVKEYs link table from WRDS, yielding a final sample of 100,570 unique firm-quarter observations for analysis (setting aside the initial training sample in 2008). Figure~\ref{fig:statistics} summarizes the dataset: event counts grow from approximately 4,500 per year in 2008 to approximately 8,000 in 2023, with roughly 2,200 distinct stocks annually. The figure also shows the character counts on average for press releases and conference calls over time.  

\section{Methodology}
\subsection{Embedding and Soft Construction}\label{sec:embedding-soft}

Bi-directional Encoder Representations from Transformers (BERT) \cite{bert} is a transformer-based language model pre-trained on large text corpora and adaptable to downstream tasks through fine-tuning. We use FinBERT \cite{finbert}, finetuned on financial text sentiment, to extract contextual embeddings for each article following \cite{earnings1}. Specifically, we represent each article as a 768-dimensional vector by applying mean pooling to the last hidden states of the first 512 tokens. 
Let $X_{c,\tau}^{k}$ denote the vectorized press release for company $c$ on announcement day $\tau$, where $k\in\bs{\text{CC, PR}}$ denote the text type. Let $\text{Ret}_{c,\tau}$ be the corresponding return. We use Ridge regression with weights $w$ to generate return scores (soft information) following \cite{ckx}:
\begin{equation}
  \arg\min_w \|X_{c,\tau}^k w-\text{Ret}_{c,\tau}\|_2^2 + \lambda \|w\|_2^2,
\end{equation}
where $\lambda$ is chosen from $[10^{-5}, 10^{-4},...,10^2]$ using cross validation with splits ensuring the training and test sets respecting the temporal order of data.
We adopt a rolling window framework as in \cite{earnings1}: the model is trained on data from year $t$ and then used to generate return scores $\hat{\text{Ret}}_{c,\tau}^k=X_{c,\tau}^k w$ (denoted $\text{Soft}_{c,\tau}^k$) for year $t+1$. This process is repeated annually from 2008 to 2023.


\subsection{Alignment with Optimal Transport}

The basic embeddings of the first 512 tokens give us a general idea on what the PR and CCs cover.\footnote{We can instead average over all PR paragraphs or Q\&A pairs, but that degrades the prediction power.} However, it is hard to disentangle the parts of the PR or CC that has the most prediction power. To this end, we apply the idea of optimal transport to find the common or distinct parts of PR and CC.

Optimal transport (OT) \cite{monge1781, kantorovich1942, Santambrogio2015} provides a principled way to map probability mass from one distribution to another while minimizing transport cost. 
Prior work shows that OT produces self-normalized and sparse yet semantically meaningful correspondences by jointly considering all pairwise relationships rather than selecting matches independently \cite{ot-alignment, swanson2020}. 
In contrast, greedy matching based solely on maximum cosine similarity may fail to capture situations where multiple paragraphs in one document correspond to the same discussion point in the other document. 
This feature is particularly important in conference calls, where several analyst questions may revisit or elaborate on the same information disclosed in the press release. 
By enforcing a globally consistent transport plan under probability mass constraints, OT allows semantically related paragraphs to share alignment mass while suppressing weak matches. As a result, the OT alignments are both more stable and more interpretable than raw cosine similarities.

There are two versions of OT optimization problem. The original Monge formulation \cite{monge1781} is often intractable and may not even admit a solution unless restrictive conditions are met \cite{cfm2002}. Instead, we adopt the Kantorovich formulation \cite{kantorovich1942}, and in particular its discrete version as described in \cite{cuturi2020}. Let $\bs{x_i}_{i=1}^n$ and $\bs{y_j}_{j=1}^m$ denote two point sets with associated probability distributions $p_x$ and $p_y$, where $\sum_{i=1}^n p_{x_i}=1$ and $\sum_{j=1}^m p_{y_j}=1$. The cost of moving mass from $x_i$ to $y_j$ is given by $c_{ij}=c(x_i, y_j)$. The optimal transport problem seeks a transport plan $\gamma_{ij}\geq 0$ that minimizes the total cost of moving probability mass:
\begin{equation}
  \min_{\gamma} \bs{\sum_{ij} c_{ij} \gamma_{ij} : \gamma_{ij}\geq 0, \sum_i \gamma_{ij} = p_{y_j}, \sum_j \gamma_{ij} = p_{x_i}}
\end{equation}
In practice, the optimal $\gamma$ is approximated via entropic regularization using the Sinkhorn algorithm~\cite{sinkhorn}. A large value of $\gamma_{ij}$ indicates that $x_i$ and $y_j$ are likely to have similar semantics. We applied Sparse-Plus-Low-Rank solver in the \texttt{RegOT} package by \citet{sinkhorn-splr} to improve computational efficiency.\footnote{With multiprocessing on 20 cores, it takes less than 1 minute per year, similar to raw cosine similarity computation.} 

In this paper, we follow \citet{ot-alignment} but extend the method from word-level to paragraph-level alignments. Specifically, each pair of PR and CC, $X^{\text{PR}}$ and $X^{\text{CC}}$ (we omit the stock-day subscripts to simplify notation), are treated as two high-dimensional sets consisting of $n$ and $m$ points (paragraphs) each. Each paragraph is assigned equal probability mass. The cost matrix is defined using pairwise cosine distance, $c_{ij} = 1 - \xi_{ij}$, where $\xi_{ij} = X_{i}^{\text{PR}}\cdot X_{j}^{\text{CC}}$ is the cosine similarity between paragraph $i$ in PR and paragraph $j$ in CC, and scaled to $[0,1]$ with min-max normalization. 

We compute transport plans in both directions: PR-to-CC and CC-to-PR, denoted $\gamma$ and $\gamma'$. 
We further threshold $\gamma$ and $\gamma'$ by 95\%-tile threshold.
If $\gamma_{ij}$ falls within the top 5\% of the matrix, we mark a forward alignment, producing the binary forward alignment matrix $A\in \bs{0,1}^{n\times m}$. The backward alignment matrix $B\in \bs{0,1}^{m\times n}$ is computed analogously.
In some cases, where the managers and analysts discuss questions irrelevant to press release, OT may still produce spurious alignments, because the pairwise cosine similarities are low overall. 
To filter these out, we use the raw pairwise cosine similarity matrix, and retain only alignments where $\xi_{ij} \geq \xi_{thres}$, where $\xi_{thres}$ is a hyperparameter that is dependent on the distribution of cosine similarities of the embedding model. We choose $\xi_{thres}=0.7$ based on the median on the full dataset. 
The final alignment matrix is obtained as the intersection of three matrices: $\calA = A \odot B^T \odot (\xi_{ij}\geq \xi_{thres})$,
where $\odot$ denotes elementwise multiplication. Paragraphs $i$ in PR and $j$ in CC are aligned if $\calA_{ij}=1$. 
This bidirectional matching ensures that $i$ is matched to $j$ if and only if $j$ is one of the most-likely transport destination of $i$ in the forward direction and simultaneously, $i$ is one of the most-likely transport destination of $j$ in the reverse direction. 
Averaging paragraph-level embeddings in each group generates four other types of embeddings $k\in \bs{\text{AP, UP, AC, UC}}$, denoting the aligned (shared) / unaligned (distinct) parts of PR and CC respectively. We then refit the return-prediction models using each subset of embeddings separately, following the same specification as in Section~\ref{sec:embedding-soft}. This procedure yields four additional soft-information scores: Soft$^{\text{AP}}$ for aligned PR, Soft$^{\text{UP}}$ for unaligned PR, Soft$^{\text{AC}}$ for aligned CC, and Soft$^{\text{UC}}$ for unaligned CC.

\subsection{Topic Discovery with oLDA}
To understand what is unique in each component of PR and CC found in the alignment process, we apply online Latent Dirichlet Allocation (oLDA) \cite{olda, olda2}. LDA provides a lower-dimensional representation of the document-term matrix by modeling each article as a multinomial distribution over topics, where each topic is a probability distribution over terms. Articles are then represented as mixtures of topics, capturing underlying themes. 

To construct the topic dictionary for the earnings calls covering both PR and CC, we firstly concatenate the text from both sides and apply standard text preprocessing pipeline to remove URLs, stopwords, and numerical values. The cleaned articles are converted into a document-term matrix using a count vectorizer, then an oLDA model is incrementally trained year by year with parameters: learning offset $\tau_0=1$, decay $0.5$, document-topic prior $\alpha=1$, and topic-word prior $\eta=1$. We use the topic model trained on years $[2008, t-1]$ to vectorize the text in year $t$, minimizing the look-ahead bias in vocabulary and topic identification.
We use 50 topics with 20 key tokens each, representing each article as a 50-dimensional vector. To improve topic interpretability, we leverage GPT-4o-2024-08-06 \cite{gpt4o} for labelling topics.\footnote{More recent models such as GPT-5.4 do not allow for temperature to be 0.0, which is not preferable for reproducibility.}

\section{Empirical Analysis}

\subsection{Baseline Return Scores}

We first establish how the content of the press release and conference call embeddings explain announcement-day abnormal returns, before any alignment decomposition. To do so, we run the following cross-sectional regression over the 2009–2023 period:
\begin{equation}\label{eq:regression-basic}
\text{AR}_{c,\tau} = \alpha + \beta_0\,\text{Surprise}_{c,\tau} + \beta_1\,\text{Soft}_{c,\tau}^k + \epsilon
\end{equation}
where $\text{AR}_{c,\tau}$ is the announcement-day abnormal return for firm $c$ on date $\tau$, computed as the stock return from the 4 p.m. close preceding the announcement to the 4 p.m. close on the following day, minus the return on the size- and book-to-market-matched Fama–French portfolio \citep{martineau2021rest}. Standard errors are two-way clustered by firm and announcement date.

To measure the importance of each independent variable, we compute SHAP values \citep{shapexact, SHAP} with normalization:
$$I_j = \frac{1}{N}\sum_{i=1}^{N}\left|\,\underset{S \subset X \setminus \{j\}}{\mathbb{E}}\bigl[y_{S \cup \{j\}} - y_S\bigr]\right|, \qquad \hat{I}_j = \frac{I_j}{\sum_{k=1}^{n} I_k} \times 100,$$
where $I_j$ is the mean absolute SHAP value for feature $j$ across $N$ observations, capturing the expected change in model output when $j$ is added to an arbitrary feature subset $S \subset X \setminus \{j\}$, and $\hat{I}_j$ normalizes this importance across all $n$ features.

Table~\ref{tab:regression-basic} reports the results. Columns (2) and (3) show that both Soft$^{\text{PR}}$ and Soft$^{\text{CC}}$ individually explain announcement-day abnormal returns, with SHAP shares exceeding that of earnings surprise, indicating that soft information from either disclosure channel is more important for cross-sectional return variation than hard information alone. Column (4) includes all three predictors jointly, raising $R^2$ from 2.77\% (surprise only) to 5.33\%.\footnote{Explaining announcement day return with linear models is difficult. Future work could incorporate nonlinearities to improve out-of-sample performance.} Conference call content is the dominant soft-information signal: a one-standard-deviation increase in Soft$^{\text{CC}}$ (0.015) is associated with a 1.0\% increase in abnormal returns ($0.015 \times 0.664$), compared with 0.8\% for an equivalent move in Soft$^{\text{PR}}$ ($0.016 \times 0.485$). Both exceed the contribution of earnings surprise in terms of SHAP value.

\begin{table}[!htb]
\caption{Cross-Sectional Regression Analysis}\label{tab:regression-basic}

\centering

\resizebox*{0.8\linewidth}{!}{\input{tables/regression_basic.tex}}
\end{table}

\subsection{Alignment Return Scores}

\begin{figure}[!htb]
  \centering
  \begin{subfigure}[b]{0.48\linewidth}
  \centering
  \includegraphics[width=\linewidth]{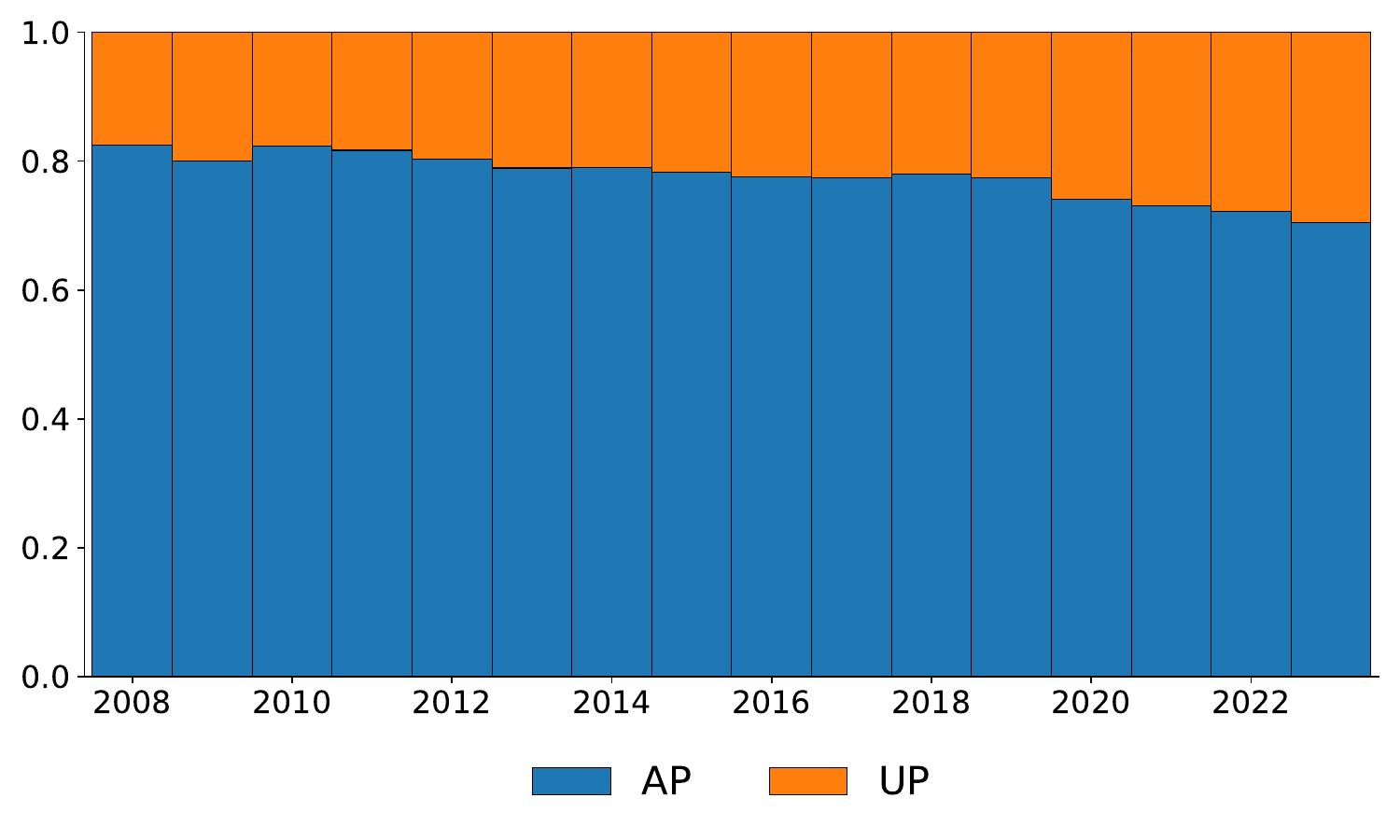}
  \caption{PR}
  \end{subfigure}
  \hfill
  \begin{subfigure}[b]{0.48\linewidth}
  \centering
  \includegraphics[width=\linewidth]{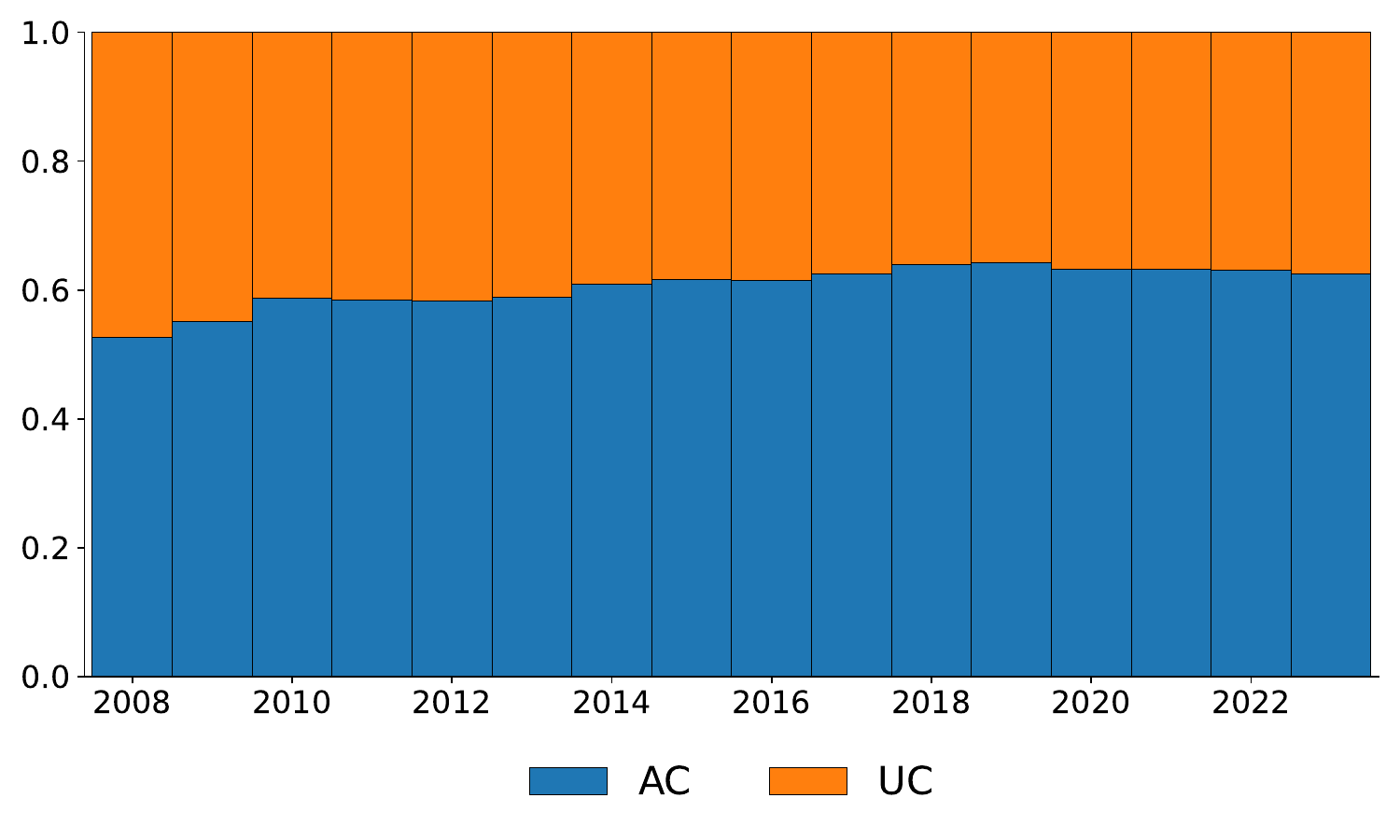}
  \caption{CC}
  \end{subfigure}
  \vspace{-0.3cm}
  \caption{Proportion of Aligned/Unaligned Paragraphs in PR and CC}\label{fig:align-proportion}
  \Description{Proportion of aligned/unaligned paragraphs in PR and CC}
\end{figure}

Figure~\ref{fig:align-proportion} shows the proportion of paragraphs between press releases and conference calls that are aligned using optimal transport. On average, approximately 80\% of press release paragraphs are semantically aligned with content discussed in the conference call, whereas only 50\%–60\% of conference call paragraphs have content that is similar in press releases. The average cosine similarity among aligned paragraph pairs is 0.86 (ranging from 0.81–0.90), compared with 0.68 (ranging from 0.56–0.78) among unaligned pairs. This suggests that nearly all of what managers put in writing is revisited during the call. Meanwhile, conference calls contain a substantial amount of unique content, reflecting the role of analyst questions in drawing out information that managers chose not to disclose in the press release. 

By contrast, greedy matching on maximum cosine similarity between paragraphs produces much fewer alignments and fails to capture the broader semantic overlap: only around 20\% of press release paragraphs and less than 10\% of conference call paragraphs are matched. Although matched pairs have mechanically higher cosine similarities under greedy (0.89, ranging from 0.84--0.93), the average cosine similarity among unmatched pairs also remains relatively high at 0.80 (ranging from 0.71--0.87), indicating that greedy matching discards many semantically related paragraphs.
Optimal transport instead accounts for the full distribution of semantic content across documents, yielding substantially greater coverage while maintaining high-quality alignments. In subsequent analyses, we therefore focus on the optimal transport approach.

We next assess whether this decomposition helps to better explain announcement day abnormal returns. We replace Equation~\eqref{eq:regression-basic} with the following specification:
\begin{align}\label{eq:regression}
	\text{AR}_{c,\tau} = \alpha &+ \beta_0 \text{Surprise}_{c,\tau} + \beta_1 \mathbb{1}_{c,\tau}^{\text{AP}} \text{Soft}_{c,\tau}^{AP} + \beta_2 \mathbb{1}_{c,\tau}^{\text{UP}} \text{Soft}_{c,\tau}^{UP} \nonumber  \\
  &+ \beta_3 \mathbb{1}_{c,\tau}^{\text{AC}} \text{Soft}_{c,\tau}^{AC} + \beta_4 \mathbb{1}_{c,\tau}^{\text{UC}} \text{Soft}_{c,\tau}^{UC} \nonumber\\
  &+ \beta_5 \mathbb{1}_{c,\tau}^{\text{AP}} + \beta_6 \mathbb{1}_{c,\tau}^{\text{UP}} + \beta_7 \mathbb{1}_{c,\tau}^{\text{AC}} + \beta_8 \mathbb{1}_{c,\tau}^{\text{UC}}+\epsilon,
\end{align}
where $\mathbb{1}_{c,\tau}^{k}$ is an indicator variable equal to one if component $k$ is present in announcement $(c,\tau)$, and zero otherwise. Accordingly, the corresponding soft score is set to zero whenever the component is absent. Including both the interaction terms and the indicator variables allows us to retain a consistent sample across specifications while separately accounting for whether a component appears and the effect of the soft score conditional on appearance. The regression is estimated after accounting for perfect multicollinearity among the indicator variables.
Table~\ref{tab:regression-alignment} reports the results, with indicator variables omitted for brevity. Three findings stand out:

First, the unaligned press release component, Soft$^{\text{UP}}$ carries a negligible SHAP share across all specifications (Columns 1 and 3), and is statistically insignificant when all four measures are included. Information that appears in the press release but is never discussed on the conference call does not move prices. This suggests that investors treat such content as less informative than the content revealed during the conference call Q\&A session.

Second, after including conference call components, the aligned press release component, Soft$^{\text{AP}}$, loses much of its explanatory power (its SHAP share falls substantially from 45.10 in Column 1 to 6.29 in Column 3). Because Soft$^{\text{AP}}$ and Soft$^{\text{AC}}$ capture semantically similar content with an average cosine similarity of 0.86, this attenuation is expected: the market-relevant information in aligned press release paragraphs is effectively subsumed by the conference call's restatement of the same material, which is potentially delivered with more context when interacted with financial analysts.

Third, and most importantly, both conference call components — the aligned portion Soft$^{\text{AC}}$ and the unaligned portion Soft$^{\text{UC}}$ — contribute significantly to the cross-sectional explanation of abnormal returns (Columns 2 and 3). The SHAP share of Soft$^{\text{AC}}$ is large, reflecting the informativeness of content shared between the two disclosures, but Soft$^{\text{UC}}$ accounts for a significant additional share. Together, these results indicate that announcement-day price formation is driven primarily by what is said on the conference call, independently of whether that content echoes the press release or goes beyond it. Information that is disclosed only in the press release and not reiterated during the Q\&A plays little role.



\begin{table}[!htb]
\caption{Alignment Regression}\label{tab:regression-alignment}

\centering 

\resizebox*{0.6\linewidth}{!}{\input{tables/regression_alignment.tex}}
\end{table}

\begin{figure}[!htb]
	\centering
  \begin{subfigure}[b]{0.48\linewidth}
  \centering
  \includegraphics[width=\linewidth]{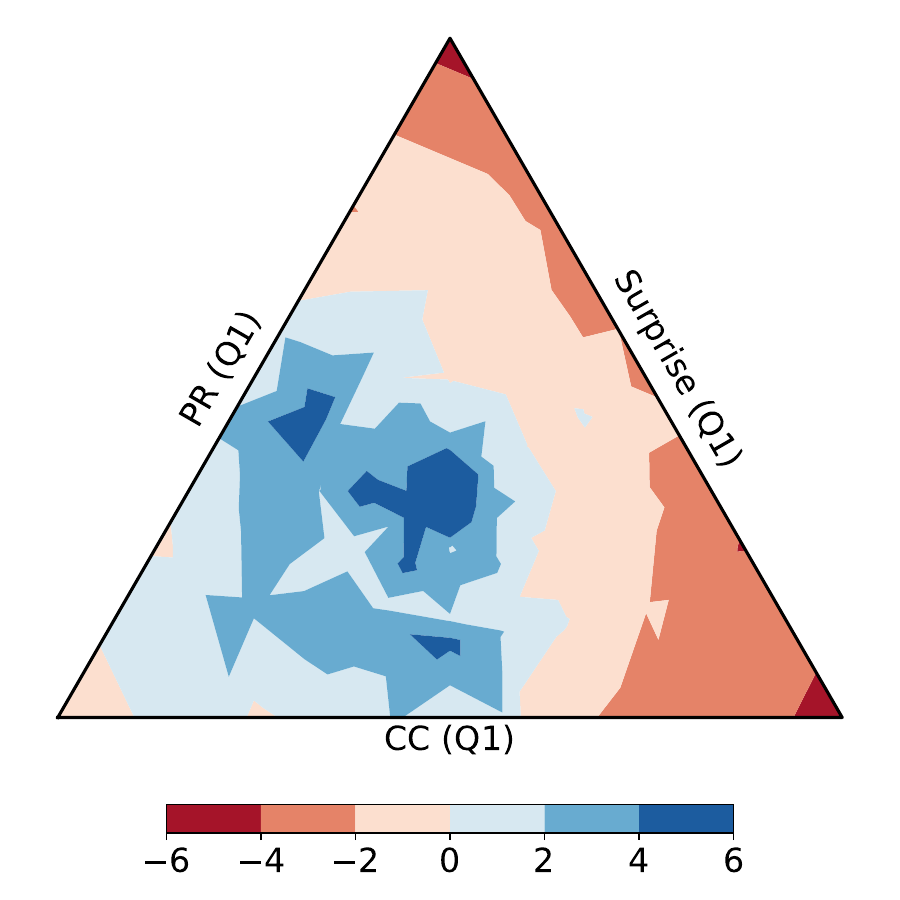}
  \caption{Basic\label{fig:avg-ret-basic}}
  \end{subfigure}
  \hfill
  \begin{subfigure}[b]{0.48\linewidth}
  \centering
  \includegraphics[width=\linewidth]{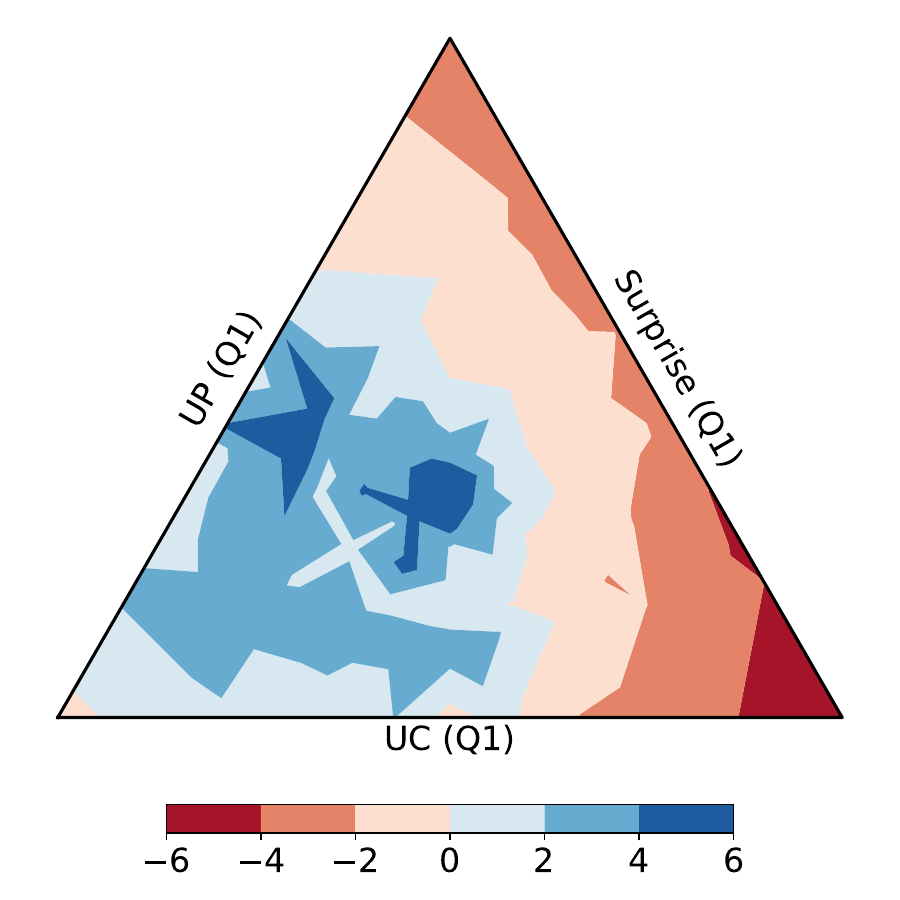}
  \caption{Alignment Based\label{fig:avg-ret-aligned}}
  \end{subfigure}
	\caption{Average Returns (\%) Sorted by Surprise and Soft}\label{fig:avg-ret}
	\Description{Average return heatmap, sorted by soft and surprise}
\end{figure}

Figure~\ref{fig:avg-ret-basic} presents average announcement-day returns from a three-way sort on earnings surprise, $\text{Soft}^{\text{PR}}$, and $\text{Soft}^{\text{CC}}$. Each variable is sorted into quintiles by quarter. We then compute the average return within each of the 125($=5\times 5\times 5$) portfolios and visualize the results using triple-sort diagrams, which highlight the cross-sectional return patterns across combinations of the three signals.

Returns are generally negative along the edges of the triangle, where at least one variable is in its lowest quintile. Towards the center, corresponding to higher quintiles across all three variables, returns become increasingly positive. The strongest variation runs from the bottom-right corner to the center. The bottom-right corner represents observations where both surprise and $\text{Soft}^{\text{CC}}$ are in their lowest quintiles (averaged over all $\text{Soft}^{\text{PR}}$ quintiles), yielding the lowest average return of $-5.3\%$. In contrast, the center corresponds to the highest quintile for all three variables and delivers the highest average return of $5.9\%$. This pattern highlights the incremental value of combining surprise with $\text{Soft}^{\text{CC}}$ in explaining announcement-day returns. The top corner, where surprise and $\text{Soft}^{\text{PR}}$ are jointly in their lowest quintiles (averaged over $\text{Soft}^{\text{CC}}$), produces a similarly negative return of $-5.2\%$.

Figure~\ref{fig:avg-ret-aligned} repeats the analysis replacing the soft-information measures with their unaligned components, UP and UC. The overall return dispersion remains similar, but portfolios formed on the lowest quintiles of UC and surprise exhibit more negative returns ($-5.6\%$) than those formed on the lowest quintiles of UP and surprise ($-4.1\%$). When all three variables are in their highest quintiles, the average return reaches $5.5\%$. These results underscore the importance of the unaligned component of conference calls, whereas the most value-relevant information from press releases appears to be concentrated in the aligned components that are reiterated during the conference call.

\subsection{Topic Analysis}
\begin{figure*}[!htb]
  \centering
  \begin{subfigure}[b]{0.48\linewidth}
  \centering
  \includegraphics[width=\linewidth]{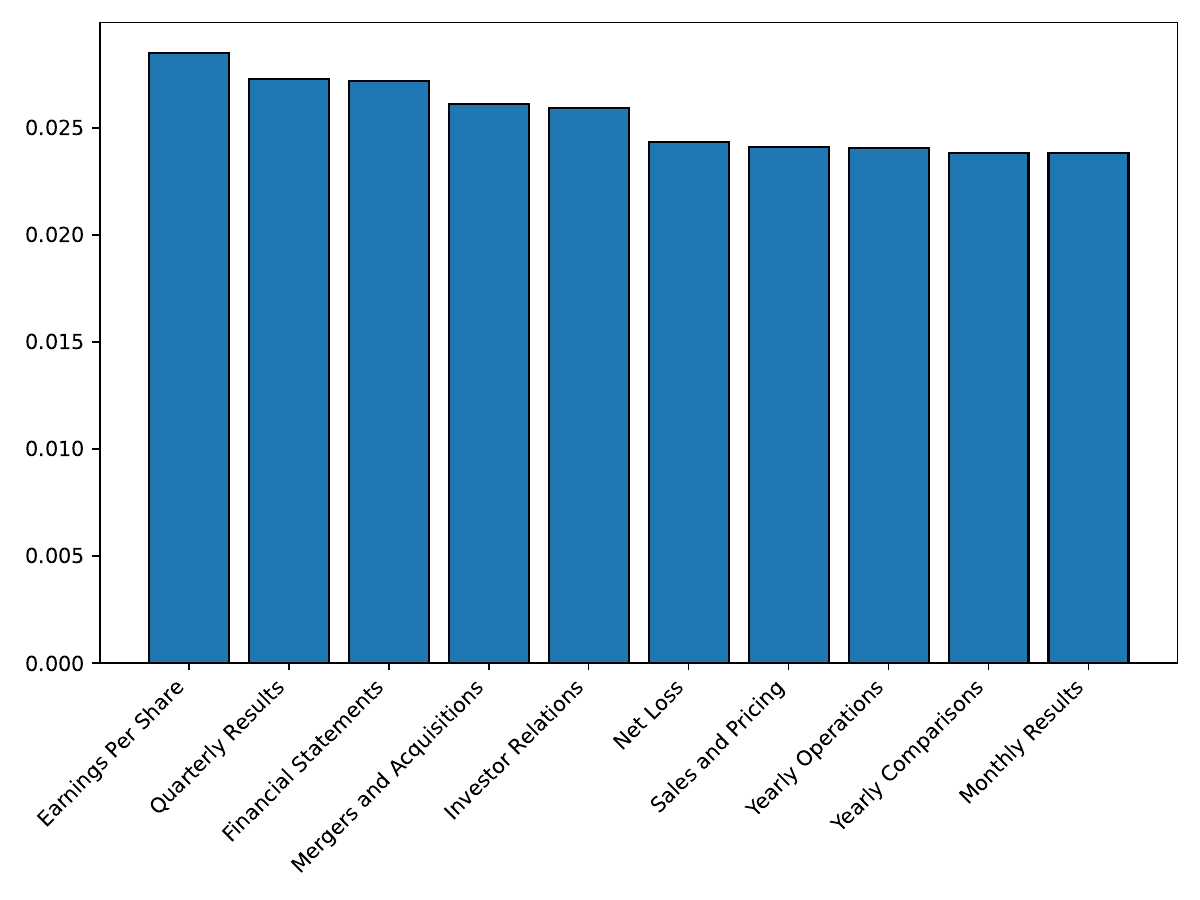}
  \caption{Topics Specific to Press Releases\label{fig:topic-analysis-up}}
  \end{subfigure}
  \hfill
  \begin{subfigure}[b]{0.48\linewidth}
  \centering
  \includegraphics[width=\linewidth]{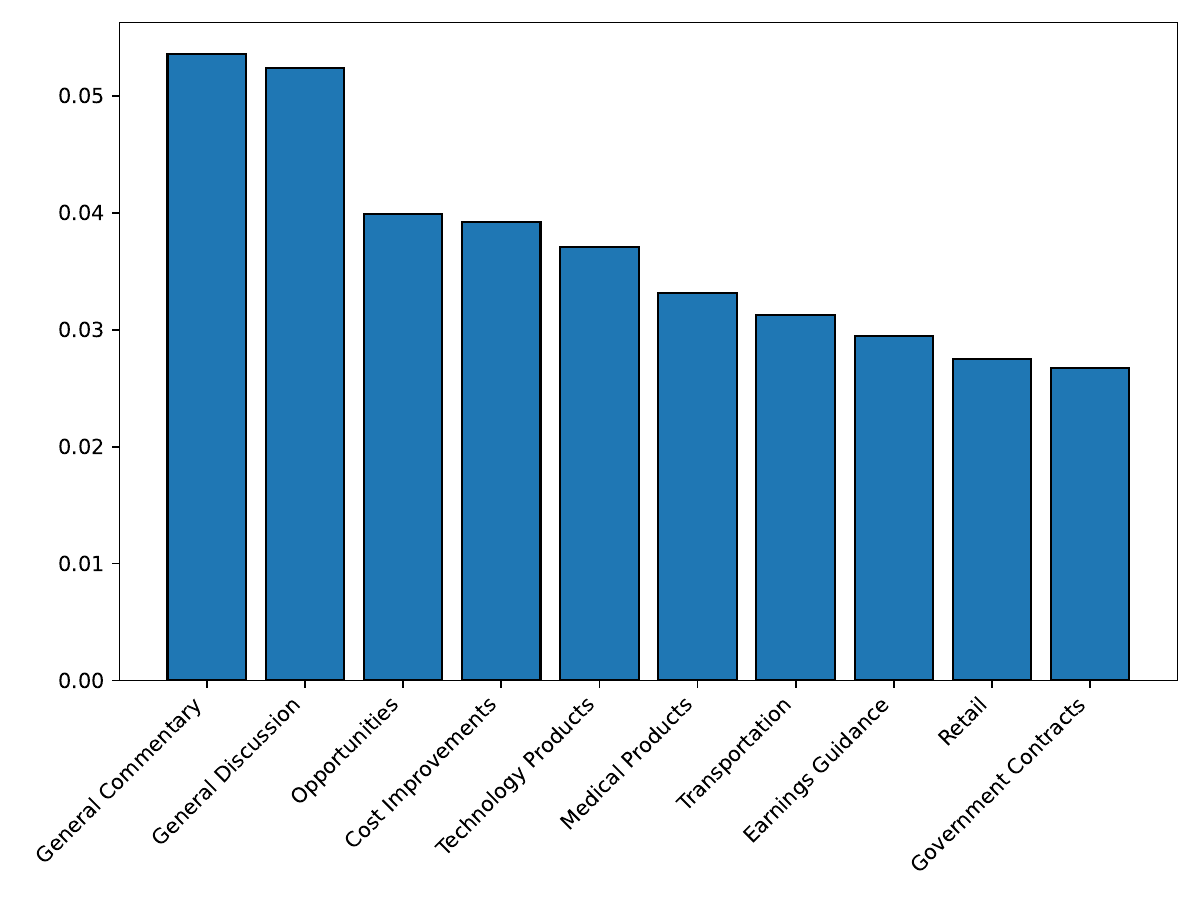}
  \caption{Topics Specific to Conference Calls\label{fig:topic-analysis-uc}}
  \end{subfigure}
  \vspace{-0.3cm}
  \caption{Topic Analysis on Unaligned Components by oLDA}\label{fig:topic-analysis}
  \Description{Topic analysis on unaligned components by oLDA}
\end{figure*}

After fitting the oLDA topic model on the full sample of concatenated PR and CC transcripts and labeling the topics using GPT-4o, we apply the resulting topic dictionary to the UP and UC components separately. Topic assignment is performed out of sample to ensure that the decomposition is not mechanically driven by in-sample fitting.
Figures~\ref{fig:topic-analysis-up} and \ref{fig:topic-analysis-uc} report the top 10 topics for UP and UC, respectively. The differences are clear. In UP, the dominant topics focus on historical performance, including discussions of past earnings, losses, and comparisons of current monthly, quarterly, or annual results with prior periods. In contrast, UC is characterized by forward-looking themes, such as growth opportunities, cost improvements, and guidance. 
Moreover, UC contains topics \textit{General Commentary} and \textit{General Discussion}, featuring less formal language used in interactive presentations and Q\&A settings. These topic patterns indicate that the model can capture discussion-type–specific terminology and distinguish retrospective reporting in UP from forward-looking and conversational content in UC. 
The following excerpt is from the UC portion of Apple’s conference call held on the evening of 2023-02-02, where the analysts and Tim Cook discussed Mac growth opportunities in a relatively informal and forward-looking manner.
\begin{center}
\begin{minipage}{0.95\linewidth}
\footnotesize

\textbf{Q:} And then just a quick follow-up on the Mac. The PC industry is expecting a decline in PC shipments this year also. How do you think about the Mac relative to kind of like where the PC industry as a whole is expecting the shipments to end up? Is there any color you can give on that?

\noindent\textbf{A:} The industry is very challenged, as you say. It's -- the industry is contracting. I think from us, though, is -- and I don't know how this year will play out, so I don't want to predict the year. But over the long run, we have a market that is a reasonable sized market, a big market. And we have low share, and we have a competitive advantage with Apple silicon. And so strategically, I think we're well positioned in the market, albeit I think it will be a little rough in the short term.
\end{minipage}
\end{center}

\subsection{Long-Run Return Prediction}

The preceding analysis focuses on announcement-day returns. A long-standing finance literature has documented the post-earnings announcement drift (PEAD) phenomenon, whereby earnings surprises (hard information) continue to predict stock returns over several weeks following the announcement \cite{ball1968,bernard1989post}. More recently, \citet{martineau2021rest, gregoire2022earnings} find that markets now efficiently incorporate earnings surprise content on announcement days, but \citet{meursault2023pead} provide suggestive evidence that the soft information in conference calls helps to predict returns over longer horizons. We next examine whether this predictability varies across the aligned and unaligned components of press releases and conference calls, and whether the decomposition reveals patterns that are otherwise unobservable if one examines the aggregate information embedded in the text data.

To capture post-announcement return dynamics, we examine Buy-and-Hold Abnormal Returns (BHAR):
$$\text{BHAR}[t_1, t_2] = \prod_{t=t_1}^{t_2}(1+\text{Ret}_{c,t}) - \prod_{t=t_1}^{t_2}(1+\text{Ret}_{p,t}),$$
where $\text{Ret}_{c,t}$ is the daily return of firm $c$ and $\text{Ret}_{p,t}$ is the return on the size- and book-to-market-matched Fama–French portfolio on day $t$. We replace the dependent variable in Equation~\ref{eq:regression} with two measures: AR[1] = BHAR[1,1] and BHAR[2,61], corresponding to the day immediately following the announcement and the subsequent three-month window, respectively.

Table~\ref{tab:regression-bhar-base} reports results from regressing AR[1] and BHAR[2,61] on earnings surprise and the aggregate soft-information measures from press releases and conference calls. Columns (1)–(2) use the full sample; Columns (3)–(4) exclude microcap stocks; Columns (5)–(6) focus on microcaps stocks (bottom quintile of the NYSE market capitalization breakpoint). It is well known that inefficiencies like underreaction to news is concentrated among small stocks \citep{hou2020replicating}. Consistent with \citet{martineau2021rest}, earnings surprise has no significant long-horizon predictive power. For the full sample (Column 2), Soft$^{\text{PR}}$ predicts price reversals ($\beta<0$) and Soft$^{\text{CC}}$ predicts price continuation ($\beta>0$). A one standard deviation increase in Soft$^{\text{PR}}$ (0.0162) and Soft$^{\text{CC}}$ (0.0153) leads to a decrease of 0.23\% and an increase of 0.21\% in the BHAR[2,61] respectively. However, this pattern is entirely driven by microcap firms (Column 6): among all-but-microcap stocks (Column 4), neither Soft$^{\text{PR}}$ nor Soft$^{\text{CC}}$ is statistically significant. For the microcap sample, a one standard deviation increase in Soft$^{\text{PR}}$ (0.0165) and Soft$^{\text{CC}}$ (0.0152) leads to a decrease of 0.59\% and an increase of 0.72\% in the BHAR[2,61] respectively. Taken at face value, these results would suggest that conference call content has no long-run return predictability for larger firms but both press release and conference call contents help predict returns for microcap stocks.


Two mechanisms may explain the overreaction to press releases and underreaction to conference call content. First, \citet{hutton2003role} argue that forward-looking statements are initially discounted as cheap talk until they are verified by subsequent outcomes, causing investors to underweight guidance revealed during the Q\&A. Second, \citet{dellavigna2009investor}'s limited attention framework suggests that the credibility of such statements is assessed only gradually. Overreaction to press releases, by contrast, likely reflects investors placing excessive weight on information that is immediately salient and requires little additional interpretation — a pattern consistent with \citet{huang2018headline}, who find that press release headlines can trigger post-announcement overreactions.

Table~\ref{tab:regression-bhar} reports the same regressions using the aligned and unaligned components separately.  On the day following the announcement, earnings surprise and both Soft$^{\text{AC}}$ and Soft$^{\text{UC}}$ are statistically significant (Column 1), highlighting that price discovery does not occur only on the day of the announcement, but also on the next day. 
We also break down the sample by firm size. For all-but-microcap stocks (Column 3), only Soft$^{\text{UC}}$ is statistically significant in explaning next-day return. For microcap stocks (Column 5), the coefficient of Soft$^{\text{UC}}$ is significant and more than doubles in magnitude compared to Column (3), while Soft$^{\text{UP}}$ and Soft$^{\text{AC}}$ are also significant in explaning next-day return.

Examining the news effect over two to 61 days post earnings (BHAR[2,61]), Column (2) reports statistically significant and negative coefficients for Soft$^{\text{AP}}$ and Soft$^{\text{UP}}$ and a statistically significant and positive coefficient for Soft$^{\text{UC}}$ for the full sample. This suggests that the reversal effect predicted by the content of press releases documented in Table~\ref{tab:regression-bhar-base} is driven by both the aligned and unaligned components of the press release, while the underreaction effect predicted by the content of conference calls is driven only by the unaligned component of the conference call.

Looking at press releases, in the full sample regression (Column 2), Soft$^{\text{AP}}$ and Soft$^{\text{UP}}$ share SHAP values of similar magnitudes (22.91 and 22.31). 
Comparing Columns (4) and (6), the predicted reversals from the aligned and unaligned content of press releases is only present for microcap stocks and not for all-but-microcap stocks. 
For microcap stocks, the SHAP value of Soft$^{\text{UP}}$ (27.48) is larger than that of Soft$^{\text{AP}}$ (18.91). A one standard deviation (0.0061) increase in Soft$^{\text{UP}}$ predicts a 0.90\% decrease in stock returns over the [2,61] periods post-earnings. 

Considering conference calls, in the full sample regression (Column 2), Soft$^{\text{UC}}$ has the largest share of SHAP value at 26.62. For all-but-microcap stocks (Column 4), only Soft$^{\text{UC}}$ helps to predict returns over the [2,61] window. A one standard deviation (0.012) increase predicts a 0.22\% continuation in stock returns. Decomposing conference call content into its aligned and unaligned parts thus uncovers a return-predictability signal that is statistically significant for larger firms that is entirely unobservable without the decomposition in Table~\ref{tab:regression-bhar-base}. For microcap stocks (Column 6), both Soft$^{\text{AC}}$ and Soft$^{\text{UC}}$ predict continuation in stock returns. A one standard deviation increase for Soft$^\text{AC}$ (0.020) and Soft$^\text{UC}$ (0.013) predicts positive returns of 0.57\% and 0.62\%, respectively.
Overall, consistent with \citet{hou2020replicating} and \citet{martineau2021rest}, most post-earnings-announcement return predictability is concentrated among microcap stocks. However, the aligned/unaligned decomposition reveals a distinct source of predictability that extends beyond microcaps: the unaligned, forward-looking component of conference calls predicts subsequent returns even for large and liquid firms.\footnote{In unreported results, greedy matching yields qualitatively similar conclusions for both announcement-day and long-run returns, although return predictability is concentrated in the unaligned components due to limited aligned portion.}

\begin{table}[!htb]
\caption{BHAR Regression}\label{tab:regression-bhar-base}

\centering 

\resizebox*{\linewidth}{!}{\input{tables/regression_bhar_base.tex}}
\end{table}

\begin{table}[!htb]
\caption{BHAR Regression (Alignment)}\label{tab:regression-bhar}

\centering 

\resizebox*{\linewidth}{!}{\input{tables/regression_bhar_alignment.tex}}
\end{table}

\section{Conclusion}
In this paper, we show that soft information from both press releases and conference calls predicts announcement-day returns, with conference call content exhibiting stronger explanatory power. Both sources of soft information outperform earnings surprise (hard information). 
Using optimal transport, we decompose press release and conference call content into aligned (shared) and unaligned (unique) components based on semantic similarity. We find that the aligned components in both disclosures significantly predict returns. More importantly, the unaligned component of conference calls is also informative, reflecting forward-looking discussions and managerial insights that are not emphasized in press releases.
Extending the analysis to post-announcement performance using BHAR, we show that markets incorporate both hard and soft information rapidly, leaving little predictability for the day immediately following the announcement. However, over longer horizons approaching the next quarterly announcement, distinct patterns emerge: investors appear to overreact to information conveyed in press releases, particularly in microcap firms, while underreacting to incremental, forward-looking information conveyed uniquely in conference calls.
Future research will investigate how aligned and unaligned disclosure components contribute to the prediction of risk factors and other sources of return variation, while leveraging more capable embedding models, refined parameter choices, and generative language models to improve the identification of semantic alignments.
We also plan to explore more flexible, potentially nonlinear modeling approaches to better capture the complexity of market reactions to different types of information.


\bibliographystyle{ACM-Reference-Format}
\bibliography{main}


\end{document}

%% file: tables/regression_basic.tex
{
\def\sym#1{\ifmmode^{#1}\else\(^{#1}\)\fi}
\begin{tabular}{@{\extracolsep{2pt}}l*{4}{c}@{}}
\hline\hline

 & (1) & (2) & (3) & (4) \\
\hline
Surprise & 0.651\sym{***} & 0.611\sym{***} & 0.623\sym{***} & 0.597\sym{***} \\
 & (0.03) & (0.03) & (0.03) & (0.03) \\
Soft$^{\text{PR}}$ &  & 0.643\sym{***} &  & 0.485\sym{***} \\
 &  & (0.03) &  & (0.03) \\
Soft$^{\text{CC-QA}}$ &  &  & 0.793\sym{***} & 0.664\sym{***} \\
 &  &  & (0.02) & (0.02) \\

\hline
$N$ & 100,570 & 100,570 & 100,570 & 100,570 \\
$R^2$ (\%) & 2.77 & 4.12 & 4.61 & 5.33 \\
SHAP(Surprise) & 100.00 & 38.54 & 34.87 & 25.88 \\
SHAP(Soft$^{\text{PR}}$) &  & 61.46 &  & 31.83 \\
SHAP(Soft$^{\text{CC-QA}}$) &  &  & 65.13 & 42.29 \\
\hline\hline
\end{tabular}
}

%% file: tables/regression_alignment.tex
{
\def\sym#1{\ifmmode^{#1}\else\(^{#1}\)\fi}
\begin{tabular}{@{\extracolsep{2pt}}l*{3}{c}@{}}
\hline\hline

 & (1) & (2) & (3) \\
\hline
Surprise & 0.629\sym{***} & 0.602\sym{***} & 0.598\sym{***} \\
 & (0.03) & (0.03) & (0.03) \\
Soft$^{\text{AP}}$ & 0.563\sym{***} &  & 0.139\sym{***} \\
 & (0.03) &  & (0.03) \\
Soft$^{\text{UP}}$ & 0.181\sym{***} &  & 0.021 \\
 & (0.05) &  & (0.05) \\
Soft$^{\text{AC}}$ &  & 0.593\sym{***} & 0.564\sym{***} \\
 &  & (0.02) & (0.02) \\
Soft$^{\text{UC}}$ &  & 0.578\sym{***} & 0.564\sym{***} \\
 &  & (0.03) & (0.03) \\

\hline
$N$ & 100,570 & 100,570 & 100,570 \\
$R^2$ (\%) & 3.39 & 5.63 & 5.66 \\
SHAP(Surprise) & 45.39 & 25.62 & 24.44 \\
SHAP(Soft$^{\text{AP}}$) & 45.10 &  & 6.29 \\
SHAP(Soft$^{\text{UP}}$) & 7.62 &  & 0.49 \\
SHAP(Soft$^{\text{AC}}$) &  & 44.09 & 40.23 \\
SHAP(Soft$^{\text{UC}}$) &  & 29.23 & 27.31 \\
\hline\hline
\end{tabular}
}

%% file: tables/regression_bhar_base.tex
{
\def\sym#1{\ifmmode^{#1}\else\(^{#1}\)\fi}
\begin{tabular}{@{\extracolsep{2pt}}l*{6}{c}@{}}
\hline\hline
& \multicolumn{2}{c}{Full sample} & \multicolumn{2}{c}{All-but-microcap} & \multicolumn{2}{c}{Microcap} \\
\cline{2-3}
\cline{4-5}
\cline{6-7}
 & AR[1] & BHAR[2,61] & AR[1] & BHAR[2,61] & AR[1] & BHAR[2,61] \\
 & (1) & (2) & (3) & (4) & (5) & (6) \\
\hline
Surprise & 0.047\sym{***} & -0.090 & 0.024 & -0.060 & 0.052\sym{***} & -0.089 \\
 & (0.01) & (0.13) & (0.02) & (0.21) & (0.01) & (0.15) \\
Soft$^{\text{PR}}$ & 0.019\sym{**} & -0.145\sym{*} & 0.014 & -0.013 & 0.036\sym{*} & -0.359\sym{**} \\
 & (0.01) & (0.08) & (0.01) & (0.06) & (0.02) & (0.18) \\
Soft$^{\text{CC}}$ & 0.043\sym{***} & 0.141\sym{**} & 0.032\sym{***} & 0.013 & 0.071\sym{***} & 0.479\sym{***} \\
 & (0.01) & (0.06) & (0.01) & (0.05) & (0.02) & (0.18) \\

\hline
$N$ & 100,570 & 100,570 & 72,162 & 72,162 & 28,408 & 28,408 \\
$R^2$ (\%) & 0.14 & 0.02 & 0.05 & 0.00 & 0.24 & 0.05 \\
SHAP(Surprise) & 33.78 & 17.52 & 13.63 & 41.17 & 44.01 & 14.63 \\
SHAP(Soft$^{\text{PR}}$) & 20.90 & 42.41 & 25.48 & 29.10 & 19.00 & 37.04 \\
SHAP(Soft$^{\text{CC-QA}}$) & 45.32 & 40.07 & 60.89 & 29.73 & 36.99 & 48.32 \\
\hline\hline
\end{tabular}
}

%% file: tables/regression_bhar_alignment.tex
{
\def\sym#1{\ifmmode^{#1}\else\(^{#1}\)\fi}
\begin{tabular}{@{\extracolsep{2pt}}l*{6}{c}@{}}
\hline\hline
& \multicolumn{2}{c}{Full sample} & \multicolumn{2}{c}{All-but-microcap} & \multicolumn{2}{c}{Microcap} \\
\cline{2-3}
\cline{4-5}
\cline{6-7}
 & AR[1] & BHAR[2,61] & AR[1] & BHAR[2,61] & AR[1] & BHAR[2,61] \\
 & (1) & (2) & (3) & (4) & (5) & (6) \\
\hline
Surprise & 0.047\sym{***} & -0.092 & 0.024 & -0.061 & 0.051\sym{***} & -0.090 \\
 & (0.01) & (0.13) & (0.02) & (0.21) & (0.01) & (0.15) \\
Soft$^{\text{AP}}$ & 0.019 & -0.249\sym{**} & 0.017 & -0.093 & 0.032 & -0.526\sym{*} \\
 & (0.01) & (0.10) & (0.01) & (0.07) & (0.03) & (0.27) \\
Soft$^{\text{UP}}$ & -0.034 & -0.462\sym{**} & -0.003 & -0.083 & -0.127\sym{**} & -1.488\sym{**} \\
 & (0.02) & (0.21) & (0.02) & (0.12) & (0.05) & (0.70) \\
Soft$^{\text{AC}}$ & 0.023\sym{***} & 0.077 & 0.010 & -0.013 & 0.050\sym{***} & 0.286\sym{**} \\
 & (0.01) & (0.06) & (0.01) & (0.05) & (0.02) & (0.13) \\
Soft$^{\text{UC}}$ & 0.043\sym{***} & 0.271\sym{***} & 0.031\sym{***} & 0.183\sym{**} & 0.071\sym{**} & 0.476\sym{**} \\
 & (0.01) & (0.08) & (0.01) & (0.08) & (0.03) & (0.20) \\

\hline
$N$ & 100,570 & 100,570 & 72,162 & 72,162 & 28,408 & 28,408 \\
$R^2$ (\%) & 0.15 & 0.06 & 0.06 & 0.03 & 0.29 & 0.13 \\
SHAP(Surprise) & 25.10 & 7.63 & 11.26 & 5.03 & 26.77 & 7.23 \\
SHAP(Soft$^{\text{AP}}$) & 11.13 & 22.91 & 19.52 & 18.68 & 7.57 & 18.91 \\
SHAP(Soft$^{\text{UP}}$) & 10.66 & 22.31 & 1.59 & 8.83 & 15.23 & 27.48 \\
SHAP(Soft$^{\text{AC}}$) & 21.50 & 11.08 & 17.34 & 4.04 & 20.26 & 17.87 \\
SHAP(Soft$^{\text{UC}}$) & 27.31 & 26.62 & 37.41 & 38.42 & 18.95 & 19.56 \\
\hline\hline
\end{tabular}
}